\documentclass[preprint,12pt]{elsarticle}

\usepackage{amssymb}
\usepackage{amsmath}
\usepackage{lineno}
\usepackage{hyperref}
\usepackage{graphicx}

\journal{Journal of Computational Physics}

\usepackage{graphicx}
\usepackage{dcolumn,natbib}
\usepackage{bm}

\usepackage[utf8]{inputenc}
\usepackage[T1]{fontenc}
\usepackage{mathptmx}
\usepackage{etoolbox}

\newcommand{\beq}{\begin{equation}}
\newcommand{\eeq}{  \end{equation}}
\newcommand{\beqa}{\begin{eqnarray}}
\newcommand{\eeqa}{  \end{eqnarray}}

\begin{document}

\begin{frontmatter}
\title{Goodbye Christoffel Symbols: A Flexible and Efficient Approach for Solving Physical Problems in Curved Spaces }
\author{Miguel A. Herrada}
\address{E.T.S.I., Depto.\ de Ingenier\'{\i}a Aeroespacial y Mec\'anica de Fluidos, Universidad de Sevilla, Camino de los Descubrimientos s/n 41092, Spain}

\date{\today}
\begin{abstract}
Traditional methods for solving physical equations in curved spaces, especially in fluid mechanics and general relativity, rely heavily on the use of Christoffel symbols. These symbols provide the necessary corrections to account for curvature in differential geometries but lead to significant computational complexity, particularly in numerical simulations. In this paper, we propose a novel, simplified approach that obviates the need for Christoffel symbols by  symbolic programming and advanced numerical methods.

Our approach is based on defining a symbolic mapping between Euclidean space and curved coordinate systems, enabling the transformation of spatial and temporal derivatives through Jacobians and their inverses. This eliminates the necessity of using Christoffel symbols for defining local bases and tensors, allowing for the direct application of physical laws in Cartesian coordinates even when solving problems in curved spaces. 

We demonstrate the robustness and flexibility of our method through several examples, including the derivation of the Navier-Stokes equations in cylindrical coordinates, the modeling of complex flows in bent cylindrical tubes, and the breakup of viscoelastic fluid threads. These examples highlight how our method simplifies the numerical formulation while maintaining accuracy and efficiency. Additionally, we explore how these advancements benefit free-surface flows, where mapping physical 3D domains to a simpler computational domain is essential for solving moving boundary problems.

\end{abstract}

\end{frontmatter}
\section{Introduction}

Christoffel symbols, fundamental artifacts in differential geometry, have been indispensable for describing curvature in manifold theory and are prominently used in general relativity to express the connection between local coordinate systems and the global curvature of spacetime \cite{wald1984}. These symbols arise naturally from the Levi-Civita connection, a unique, torsion-free connection that preserves the inner product in a Riemannian space \cite{spivak1979}. The Levi-Civita connection allows for the extension of classical operators, such as the gradient and Laplacian, into non-Euclidean geometries. In other words, this connection provides a way to compare vectors at different points on a curved manifold, which is essential for describing how spaces curve and deform in general relativity. In the context of general relativity, Christoffel symbols describe the gravitational field by representing how vectors are parallel transported in a curved spacetime, which is critical for solving Einstein's field equations \cite{misner1973gravitation}.

Beyond general relativity, Christoffel symbols also find important applications in fluid mechanics, especially when dealing with flows in curved geometries such as cylindrical or spherical coordinates \cite{batchelor2000introduction}. The Navier-Stokes equations in such systems require the inclusion of these symbols to account for the curvature of the domain, ensuring that conservation laws are properly formulated in non-Cartesian spaces \cite{anderson1995computational}. Also, the manipulation of flows with free interfaces represents a major challenge that requires sometimes mapping of the physical domain into a computational one. In particular, the need to solve equations in curved or evolving domains leads to difficulties in accurately representing the fluid interface when a interface-tracking \cite{rider1998} or  boundary-fitted coordinate\cite{thompson1982} technique is used.  These techniques excel in capturing small-scale details of the interface by aligning the computational mesh with the interface itself. This approach is highly accurate for resolving fine features with relatively few grid points \cite{Ponce2018}, but entails the drawback of complex equations due to the coordinate transformations involved. Additionally, these methods struggle with topological changes like droplet break-up or coalescence, as the mesh must continuously adapt to the changing interface geometry.

On the other hand, interface-capturing methods, such as Volume-of-Fluid (VOF)\cite{hirt1981} and level set methods \cite{chang1996}, do not require the mesh to conform to the interface. Instead, these methods implicitly represent the interface  using a scalar field. While these approaches are better suited for problems with  very complex topologies, they can suffer from numerical diffusion at the interface, which reduces accuracy in cases where surface tension is a dominant force \cite{POPINET20095838}. Furthermore, VOF methods often require mesh refinement or high-order schemes to achieve the necessary resolution at the interface.

Despite their advantages in different contexts, both interface-tracking and interface-capturing methods face limitations when applied to curved or highly deformed geometries. As previously outlined, boundary-fitted coordinate methods are highly effective for the investigation of small-scale phenomena. However, the intricate nature of the resulting equations necessitates the calculation of operators in curved space using Christoffel symbols. In this paper, we propose a novel methodology that completely avoids the use of these symbols by defining a symbolic mapping from Euclidean space to curved coordinates via a Jacobian matrix. This mapping allows us to compute spatial and temporal derivatives directly in Cartesian space, simplifying both the analytical and numerical treatment of physical equations in curved spaces.

We illustrate the flexibility and robustness of our approach with three key examples. First, we derive the Navier-Stokes equations for incompressible fluid flow in cylindrical coordinates. Second, we extend the method to handle more complex geometries, such as bent cylindrical tubes. Finally, we introduce a new example of the breakup of a viscoelastic polymeric thread, demonstrating how the method can be applied to fluid mechanics problems involving complex free surfaces and viscoelasticity. These examples show that our approach not only simplifies the mathematical formulation but also enhances computational efficiency in solving challenging physical problems.

The remainder of this paper is structured as follows: Section II details the formulation of the problem, Section III presents numerical examples, including the new viscoelastic thread breakup case, and Section IV concludes with a discussion of results and future work.

\section{Formulation of the problem}
\begin{figure}
\includegraphics[width=15cm]{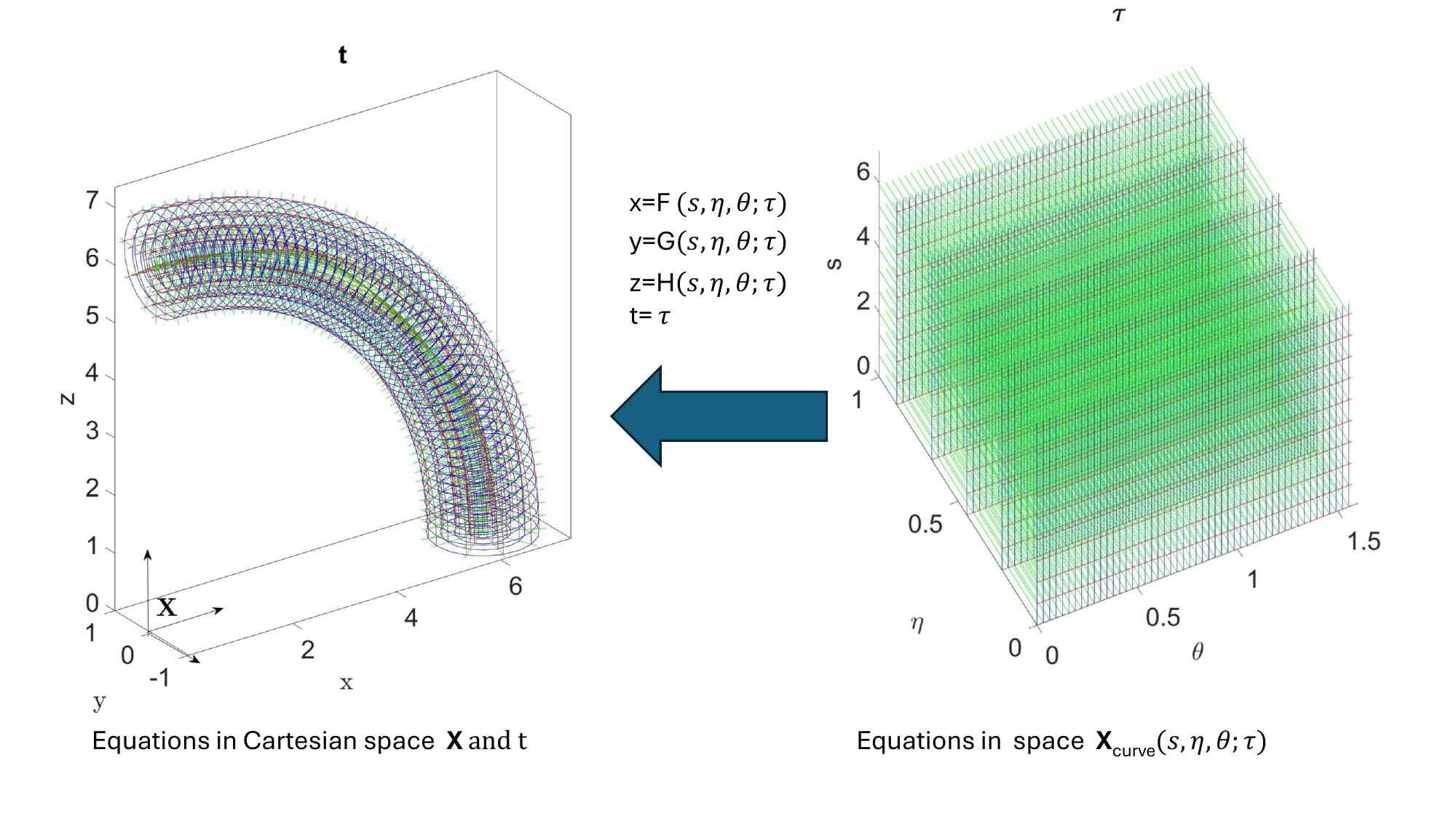}
\caption{\label{elbow0}
The problem setup involves mapping a physical 3D unsteady problem into a curve space using a non-singular continuous and derivable transformation. This allows for the solution of equations in either the physical or curved space, with the use of Christoffel symbols.\cite{aris1989vectors}}
\end{figure}

It is important to consider that our objective is to solve a physical problem in a complex domain, as illustrated in the accompanying figure \ref{elbow0}. It is assumed that the domain can be transformed by a continuous and derivable mapping to a curved coordinate space. Two approaches to solving the equations are possible. The first is to work in the original Cartesian space, where the equations are simple but the characterisation of the domain boundaries is complicated. The second is to work in the transformed space, where the equations are more complicated but the limits are characterised in a simple way. If the second option is selected, it is necessary to calculate the terms associated with the curvature, which was traditionally done with the help of Christoffel's symbols. Here we propose an alternative that avoids its use and makes the resolution of the problem more flexible. The steps to follow are

\begin{enumerate}

\item	Define a symbolic mapping   from the  three-dimensional Cartesian Euclidean space $\mathbf{X}=[x,y,z]=x\mathbf{e}_1+y\mathbf{e}_2+z\mathbf{e}_3$ and temporal domain $t$ into a curved spatio-temporal characterized by a four-dimensional representation $\mathbf{X}_{curve}=[s,\eta,\theta;\tau]$ where 
$x=F(s,\eta,\theta;\tau)$, $y=G(s,\eta,\theta;\tau)$, $z=H(s,\eta,\theta;\tau)$ are the meshlines and $t=\tau$ being  these functions known and analytic, or obtained as part of the solution.

\item	Calculate the spatial  $3\times 3$ Jacobian matrix of the transformation $\mathbf{J}_{\mathbf{X}}(s,\eta,\theta;\tau)$
\[
\mathbf{J}_{\mathbf{X}} = \begin{pmatrix}
\frac{\partial F}{\partial s} & \frac{\partial F}{\partial \eta} & \frac{\partial F}{\partial \theta} \\
\frac{\partial G}{\partial s} & \frac{\partial G}{\partial \eta} & \frac{\partial G}{\partial \theta} \\
\frac{\partial H}{\partial s} & \frac{\partial H}{\partial \eta} & \frac{\partial H}{\partial \theta}
\end{pmatrix}.
\]

and its inverse $(\mathbf{J}_{\mathbf{X}})^{-1}$  in $\mathbf{X}$. 

\item	Calculate the time derivative $3\times 1$ vector of the mapping $\mathbf{v}_t=[\frac{d s}{dt}$, $\frac{d \eta}{dt}$, $\frac{d \theta}{dt}]$ by solving the following linear system  linear equations: 

\[
 \begin{pmatrix}
\frac{\partial F}{\partial s} & \frac{\partial F}{\partial \eta} & \frac{\partial F}{\partial \theta} \\
\frac{\partial G}{\partial s} & \frac{\partial G}{\partial \eta} & \frac{\partial G}{\partial \theta} \\
\frac{\partial H}{\partial s} & \frac{\partial H}{\partial \eta} & \frac{\partial H}{\partial \theta}
\end{pmatrix}. \begin{pmatrix}  \frac{d s}{dt}\\ \frac{d \eta}{dt} \\ \frac{d \theta}{dt}\end{pmatrix}=\begin{pmatrix}  \frac{\partial F}{\partial\tau}\\ \frac{\partial G}{\partial \tau} \\ \frac{ \partial H}{\partial \tau}\end{pmatrix}.
\]

The natural basis of vectors ($\mathbf{b}_1$, $\mathbf{b}_2$, $\mathbf{b}_3$) in Euclidean space  defined as
$$
\mathbf{b}_1=\frac{\partial\mathbf{X}}{\partial s},\quad \mathbf{b}_2=\frac{\partial\mathbf{X}}{\partial \eta},\quad \mathbf{b}_3=\frac{\partial\mathbf{X}}{\partial \theta},
$$
and its associated metric tensor $\mathbf{g}$, were $g_{ij}=\mathbf{b}_i\cdot \mathbf{b}_j$, ($i=1,3$, 
 $j=1,3$) are introduced. This basis of vectors is characterized by the fact that they are tangent  to the main directions of   the mesh.  In general, the basis vectors are neither unit vectors nor orthogonal to each other. However, they are  linearly independent if the metric is not singular. Then, a vector \(\mathbf{v}\) can be expressed as:
\[
\mathbf{v} = v^k \mathbf{b_k}
\]
The components \(v^k\) are called the contravariant components of the vector \(\mathbf{v}\).

The reciprocal basis \((\mathbf{b^1}, \mathbf{b^2}, \mathbf{b^3})\) is defined by the relation:
\[
\mathbf{b_i} \cdot \mathbf{b^j} = \delta_i^j
\]
where \(\delta_i^j\) is the Kronecker delta. The vector \(\mathbf{v}\) can also be expressed in terms of the reciprocal basis:
\[
\mathbf{v} = v_k \mathbf{b^k}
\]
where \(v_k\) are the covariant components of the vector \(\mathbf{v}\).  A second-order tensor $\mathbf{S}$ can be also be represented in Euclidean space $\mathbf{X}$ in different ways depending on the basis chosen for its representation
\[
\mathbf{S}=S^{ij}\mathbf{b}_{i}\bigotimes\mathbf{b}_{j}
=S_{ij} \mathbf{b}^{i}\bigotimes\mathbf{b}^{j}=S_i^j \mathbf{b}^{i}\bigotimes\mathbf{b}_{j}=S
^i_j \mathbf{b}_{i}\bigotimes\mathbf{b}^{j}=S_{\mathbf{X} ij}\mathbf{e}_{i}\bigotimes\mathbf{e}_{j}.
\]
The fact that both the basis $\mathbf{b}_i$ and its reciprocal $\mathbf{b}^i$ are locally variable make it very difficult to extend  classical operators used in physical problems, such as the gradient or the Laplacian operator, to curved spaces based on any of these local bases. In order to address this challenge, Christoffel symbols of the first and second kind were proposed. However, the resulting expressions in curved spaces remain intricate, and the selection of appropriate Christoffel symbols remains contingent on the specific basis employed.

	In order to circumvent these complications, a methodology is put forth that obviates the necessity for employing the formulation in terms of Christoffel symbols. Generally, the velocity field is expressed in a specific basis, namely the covariant or contravariant basis or the Cartesian basis, contingent on the characteristics of the problem under consideration.  The same is true for the second-order tensors. The basis is selected with the objective of simplifying the problem or ensuring the stability of the numerical method. In our method, it is not necessary for the velocity and tensor fields to be defined with the same basis; rather, they are all projected to the Euclidean 3D Cartesian space, where all the spatial and temporal operators (gradient, Laplacian, vector product by gradient) are calculated by applying the chain rule (in a symbolic way that does not lead to  errors) and all the necessary equations are written in $X$ and $t$. 

\item	In certain cases, it is possible to reduce or simplify the number of equations and derivatives by projecting the equations into a suitable space. These equations are the ones that are then solved.

\item	The symbolic equations are converted and stored as readable functions for use in programming languages such as Python or MATLAB.  The appropriate spatial and temporal discretisation is then applied, resulting in a discrete nonlinear system to be solved.  In this article, we have chosen to use the method propose by Herrada and Montanero \cite{HM16a,JAM}. Initially used to examine flows with free interface, this approach not only takes advantage of the stored analytic functions, but also uses the analytic Jacobians of these functions to construct the numerical Jacobian and solve the system using Newton's method.

\end{enumerate}
\section{ Examples}

\subsection{The  Navier-Stokes equations for an incompressible fluid in cylindrical coordinates.}

The following example demonstrates the derivation of the Navier-Stokes equations for an incompressible fluid in cylindrical coordinates using the proposed method. The objective is to illustrate the procedure whereby these equations are derived by selecting an appropriate velocity basis and projecting the three-dimensional momentum equations using that basis.  
We begin by writing in Cartesian X-coordinates the three-dimensional equations for a Newtonian, incompressible fluid
\begin{align}
& \frac{\partial v_x}{\partial x} + \frac{\partial v_y}{\partial y} + \frac{\partial v_z}{\partial z} = 0. \label{continuity}\\
& \rho \left( \frac{\partial v_x}{\partial t} + v_x \frac{\partial v_x}{\partial x} + v_y \frac{\partial v_x}{\partial y} + v_z \frac{\partial v_x}{\partial z} \right) = -\frac{\partial p}{\partial x} + \mu \left( \frac{\partial^2 v_x}{\partial x^2} + \frac{\partial^2 v_x}{\partial y^2} + \frac{\partial^2 v_x}{\partial z^2} \right), \label{M1}\\
& \rho \left( \frac{\partial v_y}{\partial t} + v_x \frac{\partial v_y}{\partial x} + v_y \frac{\partial v_y}{\partial y} + v_z \frac{\partial v_y}{\partial z} \right) = -\frac{\partial p}{\partial y} + \mu \left( \frac{\partial^2 v_y}{\partial x^2} + \frac{\partial^2 v_y}{\partial y^2} + \frac{\partial^2 v_y}{\partial z^2} \right), \label{M2}\\
& \rho \left( \frac{\partial v_z}{\partial t} + v_x \frac{\partial v_z}{\partial x} + v_y \frac{\partial v_z}{\partial y} + v_z \frac{\partial v_z}{\partial z} \right) = -\frac{\partial p}{\partial z} + \mu \left( \frac{\partial^2 v_z}{\partial x^2} + \frac{\partial^2 v_z}{\partial y^2} + \frac{\partial^2 v_z}{\partial z^2} \right).\label{M3}
\end{align}
where $\mathbf{V}=[v_x,v_y,v_z]=v_x(x,y,z,t)\mathbf{e}_x+v_y(x,y,z,t)\mathbf{e}_y+v_z(x,y,z,t)\mathbf{e}_z$ is the velocity field in $\mathbf{X}$, $p$ is the pressure and $\rho$ and $\mu$ are the density and viscosity of the fluid. The equations (\ref{continuity})-(\ref{M3}) can be expressed more compactly using the divergence ( $\nabla \cdot$) gradient ($\nabla$),  Laplacian ($\nabla^2$) and  $(\mathbf{V} \cdot \nabla)$ operators which in Cartesian coordinates have very simple expressions
\begin{align}
 \nabla \cdot \mathbf{V} &= 0, \label{continuity1} \\
 \mathbf{M}_{\mathbf{X}}  =\rho \left( \frac{\partial \mathbf{V}}{\partial t} + (\mathbf{V} \cdot \nabla) \mathbf{V} \right) +\nabla p - \mu \nabla^2 \mathbf{V}&=\mathbf{0}.\label{MF}
\end{align}
Note that  in (\ref{MF}),  $\mathbf{M}_{\mathbf{X}}$ is a 3x1 vector equation in $\mathbf{X}$
\beq
\mathbf{M}_{\mathbf{X}}=[M_x,M_y,M_z]=M_x\mathbf{e}_x+M_y\mathbf{e}_y+M_z\mathbf{e}_z=\mathbf{0},\label{fullME}
\eeq
 that shall be solved as is or projected in appropriate directions to simplify the problem.

For a variety of problems, such as the flow in a cylindrical pipe or in a liquid jet,  cylindrical coordinates are used by aligning the axis of the pipe (or jet) with the axis of the coordinate system. The correspondence with Euclidean space is given by
$\mathbf{X}=[x,y,z]=[F,G,H]$ with $x=F(r,z_1,\theta)=r\cos\theta$, $y=G(r,z_1,\theta)=r\sin\theta$, $z=H(r,z_1,\theta)=z_1$ while $t=\tau$ is the correspondence in time. The Jacobian matrix and it inverse are given by

\[
\mathbf{J}_\mathbf{X} = \begin{pmatrix}
 cos\theta &0 & -r\sin \theta \\
\sin\theta  &0  & r\cos\theta \\
0 & 1 & 0
\end{pmatrix},\quad \mathbf{J}_\mathbf{X}^{-1} = \begin{pmatrix}
 cos\theta &\sin \theta &  1 \\
0 &0  & 1 \\
 -\frac{1}{r}\sin \theta &  -\frac{1}{r}\cos \theta & 0
\end{pmatrix},
\]

while the vector of the time derivative of the mapping is zero since the reference frame is not time-dependent,  $\mathbf{v}_t=[\frac{d r}{dt}$, $\frac{d z_1}{dt}$, $\frac{d \theta}{dt}]=[0,0,0]$.

The natutal vector basis parallel to the principal directions $r$, $z_1$, and $\theta$ ($\mathbf{b}_i$ $i=1,3$) is given by  
\beq
\mathbf{b}_1=\mathbf{e}_r=\frac{\partial \mathbf{X}}{\partial r},\quad \mathbf{b}_2=\mathbf{e}_{z_1}=\frac{\partial \mathbf{X}}{\partial z_1},\quad \mathbf{b}_3=r\mathbf{e}_\theta=\frac{\partial \mathbf{X}}{\partial \theta},\label{basis}
\eeq
where $\mathbf{e}_r$, $\mathbf{e}_{z1}$ and $\mathbf{e}_\theta$ are the vectors that form the unitary and orthogonal basis of the cylindrical coordinate system
\beq
\mathbf{e}_r(\theta)=[\cos\theta, sin\theta,0],\quad \mathbf{e}_\theta(\theta)=[-sin\theta,cos\theta,0], \quad \mathbf{e}_{z1}=[0,0,1].  \label{ecylindrical}
\eeq

The velocity field is expressed in terms of this normalised basis 
\beq
\mathbf{V}(r,z_1,\theta;\tau)=u(r,z_1,\theta;\tau)\mathbf{e}_r(\theta)+w(r,z_1,\theta;\tau)\mathbf{e}_{z1}+v(r,z_1,\theta;\tau)\mathbf{e}_\theta(\theta), \label{cilindricals}
\eeq
where $u$, $w$ and $v$ are the radial, axial and azimuthal components of the velocity field in cylindrical coordinates. Then, we calculate the partial derivatives of the velocity and pressure field with respect to the three main directions of Euclidean space (x, y, and z) and the time derivatives by applying the chain rule of differentiation. For example, 
\beq
\begin{aligned}
 \frac{\partial \mathbf{V}}{\partial x} &=& \mathbf{J}^{-1}_\mathbf{X} (1,1)\frac{\partial \mathbf{V}}{\partial r}+ \mathbf{J}^{-1}_\mathbf{X} (2,1)\frac{\partial \mathbf{V}}{\partial z_1}+\mathbf{J}^{-1}_\mathbf{X} (3,1)\frac{\partial \mathbf{V}}{\partial \theta},\\
  \frac{\partial \mathbf{V}}{\partial y} &=&\mathbf{J}^{-1}_\mathbf{X} (1,2)\frac{\partial \mathbf{V}}{\partial r}+ \mathbf{J}^{-1}_\mathbf{X} (2,2)\frac{\partial \mathbf{V}}{\partial z_1}+\mathbf{J}^{-1}_\mathbf{X} (3,2)\frac{\partial \mathbf{V}}{\partial \theta},\\
   \frac{\partial \mathbf{V}}{\partial z} &=&\mathbf{J}^{-1}_\mathbf{X} (1,3)\frac{\partial \mathbf{V}}{\partial r}+ \mathbf{J}^{-1}_\mathbf{X} (2,3)\frac{\partial \mathbf{V}}{\partial z_1}+\mathbf{J}^{-1}_\mathbf{X} (3,3)\frac{\partial \mathbf{V}}{\partial \theta},\\\label{chain}
  \frac{\partial \mathbf{V}}{\partial t} &=&\frac{\partial \mathbf{V}}{\partial \tau}+\mathbf{v}_t(1)\frac{\partial \mathbf{V}}{\partial r}+\mathbf{v}_t(2)\frac{\partial \mathbf{V}}{\partial z_1}+\mathbf{v}_t(3)\frac{\partial \mathbf{V}}{\partial \theta}.
 \end{aligned} 
\eeq
For the calculation of second derivatives, simply repeat the process described above.

Finally, substituting the velocity field given by the equation (\ref{cilindricals}), the pressure and their derivatives in  $\mathbf{X}$ and $t$, in equations (\ref{continuity1}) and (\ref{fullME}) we arrive at a system of four equations (continuity, x-momentum, y-momentum,z-momentum) that allow us to calculate $u$, $w$, $v$ and $p$ in the space $[r,z_1\theta, \tau]$. If we are solving a purely three-dimensional problem, we would simply discretise the space $[r,z_1,\theta, \tau]$ and solve these equations for each instant of time.  However, in this example our aim is to derive the Navier-Stokes equations for an incompressible fluid in cylindrical coordinates, and for this we would only have to multiply scalarly our vector equation (7) by  $\mathbf{e}_r$, $\mathbf{e}_{z1}$ and $\mathbf{e}_\theta$ and simplify the result. The momentum equations in cylindrical coordinates are 

\begin{align}
M_r &=M_\mathbf{X}\cdot \mathbf{e}_{r}=0, \text{ (radial momentum equation)} \label{Mr}\\ 
M_{z_1}&=M_\mathbf{X}\cdot \mathbf{e}_{z_1}=0,\text{ (axial momentum equation)}   \label{Mz} \\
M_\theta&=M_\mathbf{X}\cdot \mathbf{e}_{\theta}=0. \text{ (azimuthal momentum equation)}\label{Mt}
\end{align}

In regard to the continuity equation, no action is required, given that when the chain rule is applied to the spatial derivatives of the velocity, the expression in cylindrical coordinates is obtained directly. As can be seen from the procedure, the only action required is to repeatedly apply the derivation rule. To avoid errors, it is advisable to use software with symbolic packages. To illustrate the advantages of the symbolic tools in this method, we provide as supplementary material the code of the present example written in the software MATHEMATICA. For comparison, the  derivation with Christoffel symbols and the metric tensor $\mathbf{g}$ of the aforementioned equations can be found in appendix A. 

It should be noted that the vector field $\mathbf{V}$, expressed in either cylindrical or Cartesian form, can be solved for the momentum equations in either Cartesian (equation (\ref{fullME})) or cylindrical (equations (\ref{Mr})-(\ref{Mt})) bases, provided that the bases are formed by linearly independent vectors. The choice of our basis depends on the specific problem under consideration.  For example, if we know that the flow is axisymmetric ($\frac{\partial}{\partial\theta}=0$) and without swirl($v=0$), we can express the velocity field in cylindrical coordinates  to reduce the 3D problem to a 2D problem by solving the equations (\ref{continuity1}), (\ref{Mr}) and (\ref{Mz}) in two-dimensional space ($r, z_1$).

\subsection{Flow in a  Bent Cylindrical Tube}
\begin{figure}
\includegraphics[width=12 cm]{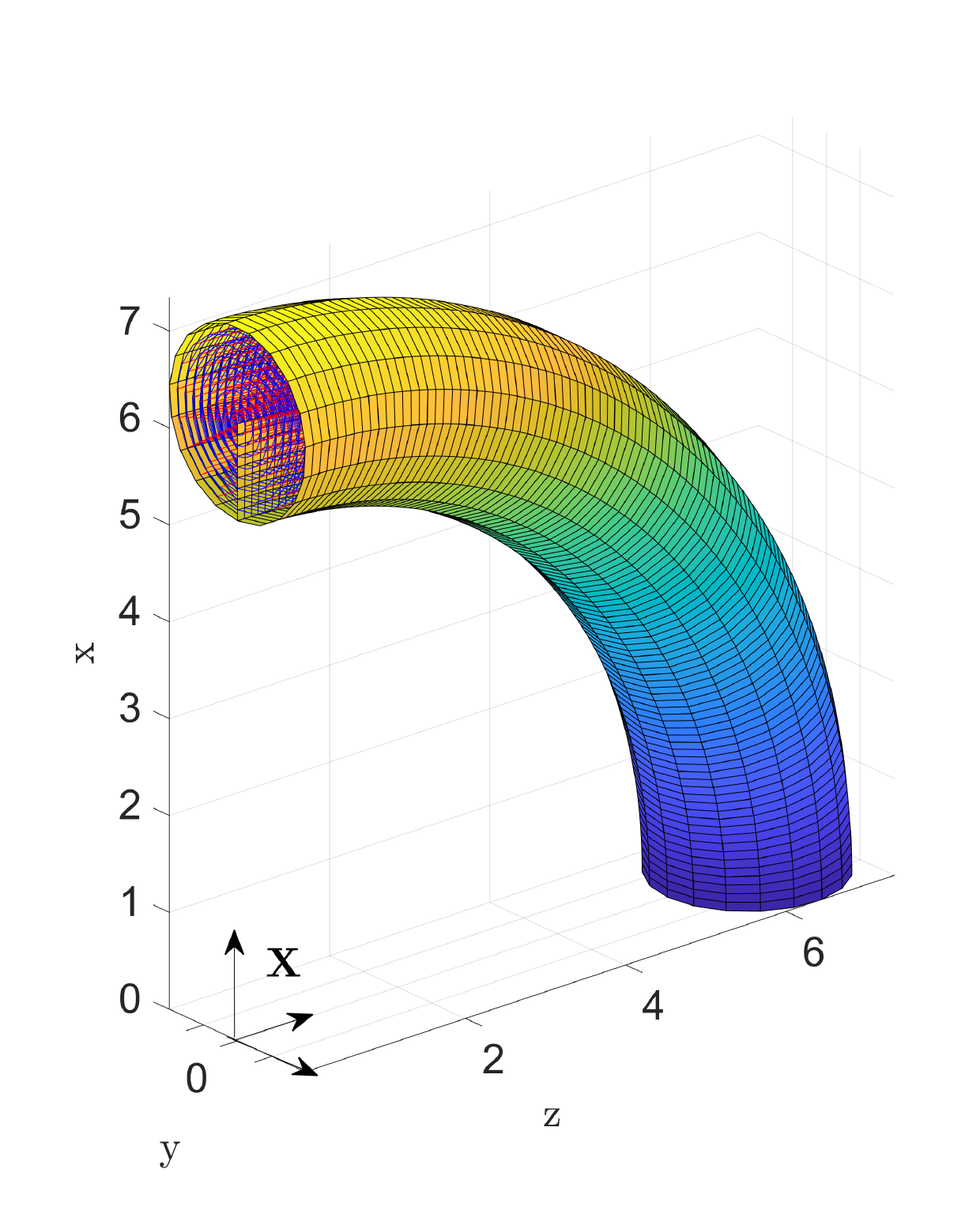}
\caption{\label{elbow}
Sketch of the flow geometry considered in this example: a cylindrical tube bent into a semicircle.}
\end{figure}

In the following example we apply the method to obtain the pressure and velocity field of Newtonian and incompressible flow in a bent cylindrical  pipe. First, we aim to parametrize the volume of a cylindrical tube with radius \( R=1 \) and length \( L \) bent into a semicircle (see figure \ref{elbow}). This implies that we will be working in a dimensionless physical space where all lengths have been rescaled with the radius of the tube $R$.

The centerline of the tube forms a semicircle in the \( x \)-\( z \) plane and extends linearly along the \( z \)-axis. The parametric equations for the centerline are:
\[
\begin{aligned}
x_c(s) &= R_c \cos s, \\
y_c(s) &= 0, \\
z_c(s) &= R_c \sin s,
\end{aligned}
\]

where \( s \) ranges from 0 to \( \pi/2 \) and \( R_c = \frac{2L}{\pi} \).

To describe the volume of the tube, we use the coordinates \(\eta \) and \( \theta \) to define a circle of radius \( 1 \) perpendicular to the centerline at each point \( s \) where \( \eta \) ranges from 0 to \( 1 \) and   \( \theta \) ranges from 0 to \( 2\pi \). Finally, the parametric equations  for the volume of the tube in the $\mathbf{X}$ space are:

\beq
\begin{aligned}
x(s, \eta, \theta)=F(s, \eta, \theta) &= \frac{\cos s \left(2L - \pi \eta \cos\theta\right)}{\pi}, \\
y(s, \eta, \theta)= G(s, \eta, \theta)&= -\eta \sin\theta, \\
z(s, \eta, \theta)=H(s, \eta, \theta) &= \frac{\sin s \left(2L - \pi \eta \cos\theta\right)}{\pi}.
\end{aligned}
\eeq

Looking for the points where the determinant of the Jacobian of the mapping becomes zero, we find that the only singular points are those on the centreline of the pipe ($\eta=0$). 

If we use mean velocity at the pipe entrance $U$, the density of the liquid $\rho$ and the radius of the pipe $R$ as characteristic magnitude the dimensionless equations for the bulk are 
\begin{align}
 \nabla \cdot \mathbf{V} &=& 0, \label{continuitynew} \\
\mathbf{M}_{\mathbf{X}} =\left( \frac{\partial \mathbf{V}}{\partial t} + (\mathbf{V} \cdot \nabla) \mathbf{V} \right) +\nabla p  -\frac{1}{Re} \nabla^2 \mathbf{V} &=&\mathbf{0}
\label{MFnew}
\end{align}

 $Re=\rho U R/\mu$ is the Reynolds number and, as in the previous example, all operators and vectors are written in the Cartesian system $\mathbf{X}$. 

For this problem we will express the velocity field in Cartesian coordinates but defined by the mapping
\beq
\mathbf{V}=v_x(s,\eta,\theta;\tau) \mathbf{e}_x+v_y(s,\eta,\theta;\tau) \mathbf{e}_y+ v_z(s,\eta,\theta;\tau) \mathbf{e}_z.\label{cartesian}
\eeq
As boundary conditions, we assume that at the entrance ($z=0$ or $s=0$) we have a Hagen-Pouiseille profile.   \beq
V_z=2(1-\eta^2),\quad V_x=0,\quad V_y=0.
\eeq
At the exit ($x=0$ or $s=\pi/2$)  the pressure is set to zero
 \beq
 p=0.
  \eeq
Finally, at the wall ($\eta=1$), the no-slip boundary conditions are considered
  \beq
  v_x=v_y=v_z=0.
  \eeq
  
As we point out before, at $\eta=0$, the mapping is singular. However, by using a Cartesian velocity field, this singularity can be circumvented by selecting a discretisation in $\eta$ that does not encompass this value, thereby obviating the necessity for additional boundary conditions. Furthermore, the selection of a Cartesian velocity field ensures that the functions $v_x$, $v_y$, $v_z$ and $p $ are periodic in the variable $\theta$. Consequently, if a Fourier discretisation is employed for the derivatives in the $\theta$ direction, it is unnecessary to impose boundary conditions in this direction.

The follow step is the same that in the previous example. We compute $\mathbf{J}_\mathbf{X}$, $\mathbf{J}^{-1}_\mathbf{X}$  and $\mathbf{v}_t$ and use them to compute the spatial and temporal derivatives in $\mathbf{X}$ and $t$  using equation (\ref{chain}).
Finally, substituting the velocity field given by the equation (\ref{cartesian}), the pressure, $p(s,\eta,\theta,\tau)$ and their derivatives in  $\mathbf{X}$ and $t$, in equations (\ref{continuitynew}) and (\ref{MFnew}) we arrive at a system of four equations (continuity, x-momentum, y-momentum, z-momentum) that allow us to calculate $v_x$, $v_y$, $v_z$ and $p$ in the bulk of or our domain.

\subsubsection{Numerical implementation}

The governing equations are integrated using a variant of the numerical method proposed by Herrada and Montanero\cite{HM16a}. A principal feature of this procedure is that the elements of the Jacobian matrix of the discretised system of equations are obtained by combining analytical functions and collocation matrices. This allows the sparsity of the resulting matrix to be exploited in order to reduce the computational time on each Newton iteration step. The numerical domain, defined as the interval $[0\leq s\leq \pi/2]\times [\epsilon\leq \eta\leq 1]\times [0\leq \theta\leq 2\pi]$, is discretised in the direction of the variable $\eta$ using $n_{\eta}$ Chebyshev spectral collocation points, while the direction of the variable $\theta$ is discretised using $n_{\theta}$ Fourier collocation points.  Conversely, in the s-direction,  forth-order finite differences with $n_s$ equally spaced points are employed.  The results presented in this example were obtained using  $\epsilon=0.0001$, $n_s=100$, $n_s=8$, and $n_\theta=8$. The temporal derivatives have been discretised using second-order backward derivatives. The method is fully implicit, so there are no stability constraints on the time step, $\Delta t$. In this example, we present only stationary solutions to the problem, and we have chosen $\Delta t=10^{10}$ to obtain such a solution using a single time step.

\subsubsection{Results}

\begin{figure}
\includegraphics[width=15 cm]{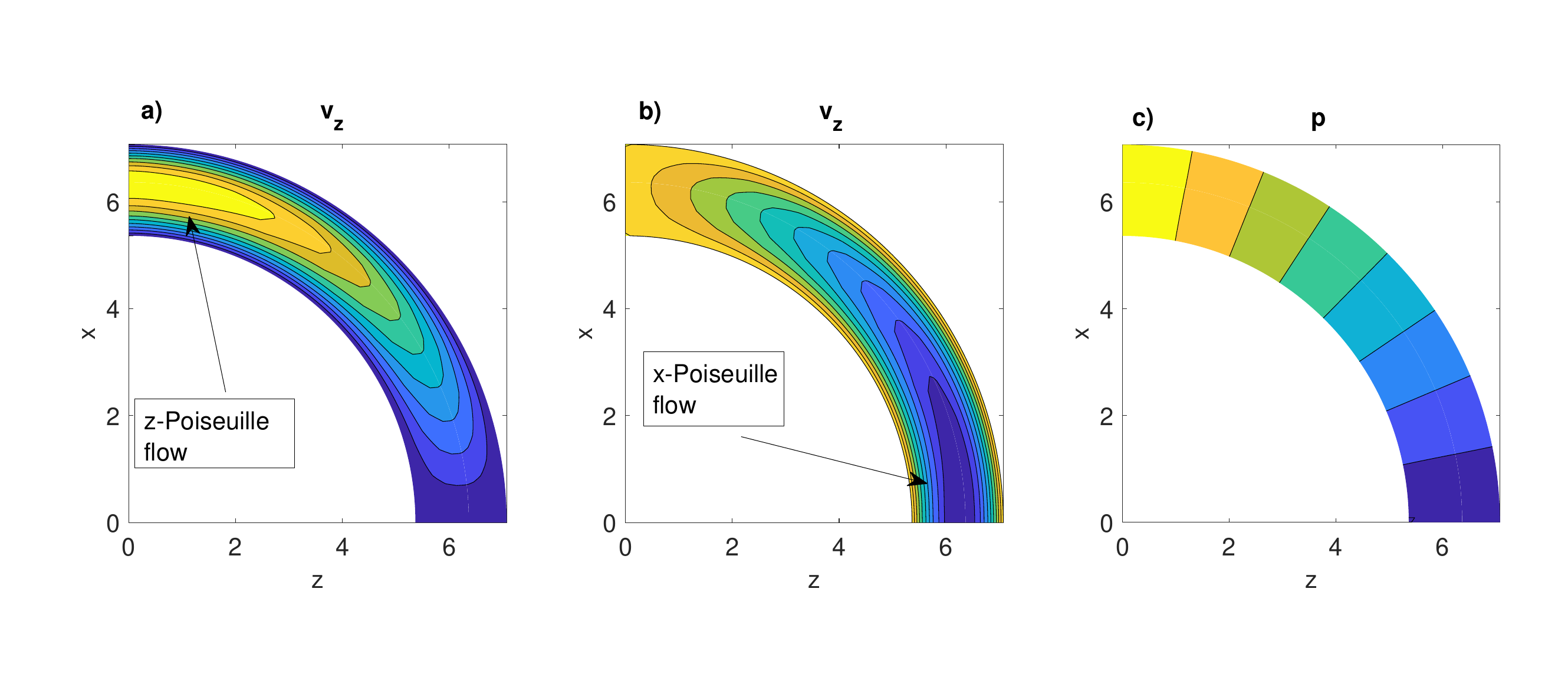}
\caption{\label{velocities}
Contours of z-velocity (a), x-velocity (b) and pressure (c) in the mid-plane of the tube ($y=0$) for $Re=0.0001$ and $L=10.$}
\end{figure}

Figure \ref{velocities} shows the results obtained for the case Re=0.0001 and $L=10$. The Reynolds number is sufficiently small, and we can confirm that a Hagen-Poiseuille profile, the elementary solution of the flow in a straight pipe, is maintained in the main direction along the length of the pipe.   Figure \ref{velocities} (c) also indicates that we have an almost unidirectional regime with a linear pressure drop along the length of the pipe.  The complete code for this example is available at

\href{https://github.com/miguelherrada/JAM/tree/main/Matlab/elbow/}{https://github.com/miguelherrada/JAM/tree/main/Matlab/elbow/}.

\subsection{Breakup of a polymeric thread as
described by the Oldroyd-B model.}
\begin{figure}
\centering
\includegraphics[width=16 cm]{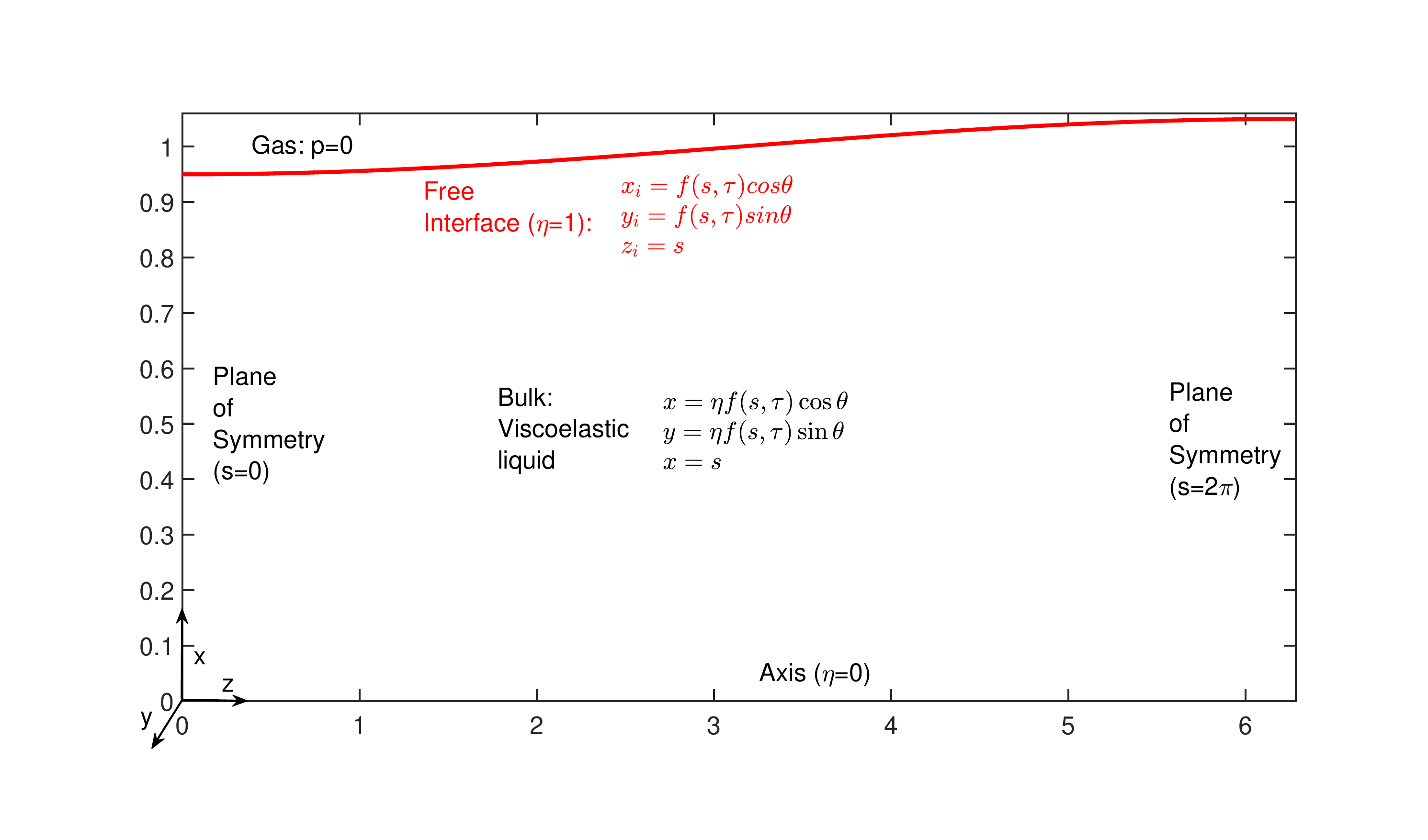}
\caption{\label{thread}
Numerical domain and boundary conditions in the viscoelastic filament problem. }
\end{figure}

The last example illustrates the breakage of a polymeric thread. We have chosen this problem because the mapping must be obtained as part of the solution, the problem is solved with second order tensors and there are surface operations, such as calculating the curvature of the interface.  An axisymmetric sinusoidal perturbation is superimposed on a cylindrical viscoelastic thread of radius R with axial wavelength $4\pi R$ and amplitude $A=0.05R$. Figure (\ref{thread})  illustrates a sketch of  this benchmark problem solved by Turkoz et al.\cite{Turkoz2018} by using
 the open source solver Basilisk developed by Stephan Popinet.\cite{POPINET2015336}  As in the previous example, we work directly in a three-dimensionless Cartesian space, $\mathbf{X}=[x,y,z]$, where we use the radius of the thread, $R$, as the characteristic length and where the symmetry of the problem is used to halve the computational domain. 
The volume of the thread  is parametrized by introducing the following mapping ($\mathbf{X}{curve}=[s,\eta,\theta;\tau]$)
\beq
x=f(s;\tau)\eta\cos\theta,\quad y=f(s,\tau)\eta\sin\theta,\quad z=s,\quad t=\tau,\label{mapping3}
\eeq
  whith  $[0\leq s\leq 2\pi]\times [0\leq \eta\leq 1] \times [0\leq \theta\leq 2\pi]$ and where function $f(s;\tau)$ must be computed as part of the solution. The surface delineating the interface with the external gas, designated as $\mathbf{X}_i=[x_i,y_i,z_i]$, is  is located at $\eta=1$.   In the reference system under consideration, the axis of symmetry is represented by the value of 0 for the parameter $\eta$.

If we use as characteristic magnitudes, the radius of thread $R$, the density $\rho$ and the capillary time $\tau_c=\sqrt{\rho R^3/\gamma}$, where $\gamma$ is the surface tension,  the three-dimensional dimensionless equations for a viscoelastic, incompressible fluid
described by the Oldroyd-B model characterized by having a solvent viscosity $\eta_s$, polymer viscosity $\eta_p$ and relaxation time $\lambda$,  are given by

\begin{align}
\nabla \cdot \mathbf{V} &=& &&0, \label{eq1}\\
 \mathbf{M}_\mathbf{X}&=&\frac{\partial \mathbf{V}}{\partial t} + (\mathbf{V} \cdot \nabla) \mathbf{V} +\nabla p -  \nabla\cdot \mathbf{\sigma}_p - S Oh  \nabla^2\mathbf{V}&=&\mathbf{0} ,\label{eq2}\\
\mathbf{C}_{\mathbf{X}\mathbf{X}}&=&\frac{\partial \mathbf{A}}{\partial t}+(\mathbf{V} \cdot \nabla) \mathbf{A} - \mathbf{A} \cdot \nabla \mathbf{V} - (\nabla \mathbf{V})^T \cdot \mathbf{A} +\frac{1}{De} (\mathbf{A} - \mathbf{I})&=&\mathbf{0}.\label{eq3}
\end{align}
where  $S=\eta_s/\eta_o$ is the solvent viscosity ratio, $Oh=\eta_0/\sqrt{R\rho\gamma}$  the Ohnesorge number and $De=\lambda/\tau_c$ the Deborah number.  Note that (\ref{eq3})  is a matrix  equation for the element $3\times 3$ element of the conformal matrix $\mathbf{A}$ in $\mathbf{X}$,
The equation relating the conformation tensor \( \mathbf{A} \) to the polymeric stress tensor \( \sigma_p \) in the Oldroyd-B model is:

\begin{equation}
\mathbf{\sigma}_p = \frac{(1-S) Oh}{De} (\mathbf{A} - \mathbf{I}),
\end{equation}
while the total stress tensor given by $\mathbf{\sigma}=\mathbf{\sigma}_p+\mathbf{\sigma}_s$, with $\mathbf{\sigma}_s=SOh \left[\nabla \mathbf{V} +(\nabla \mathbf{V})^T\right]$.  

On the other hand,  at the free surface ($\eta=1$, $\mathbf{X}=\mathbf{X}_i$), it is necessary to impose conditions of zero tangent stresses, an equilibrium of normal stresses and surface tension forces, 
\begin{equation} \label{surfaceeq}
-p-\nabla\cdot \mathbf{n}+\mathbf{n}\cdot\mathbf{\sigma}\cdot\mathbf{n}=0\quad \mathbf{t}_1\cdot\mathbf{\sigma}\cdot\mathbf{n}=0,\quad \mathbf{t}_2\cdot\mathbf{\sigma}\cdot\mathbf{n}=0,
\end{equation}
and the kinematic equation 
\beq
\frac{D \mathbf{F_{surface}}}{Dt}=0,
\eeq
where, $F_{surface}$ is the implicit equation for the interface, $\bf{n}$, and  $\bf{t}_1$ and $\bf{t}_2$  and are normal and tangential vectors to the surface and $\kappa=\nabla\cdot \mathbf{n}$ is the mean curvature. It should be noted that all operators, including those of a tensor nature, and operations between vectors, as represented in the above equations in Cartesian space, have trivial expressions.

Once the three-dimensional equations have been set out, we will simplify them, taking advantage of the fact that the problem is axisymmetric, and we will reduce their number by projecting them into the appropriate space. For this purpose we will express both the velocity field and the tensor $\mathbf{A}$ in cylindrical coordinates 
\begin{align}
\mathbf{V} &= u\mathbf{e}_r + w\mathbf{e}_{z1},\nonumber \\
\mathbf{A} &= A_{rr}\mathbf{e}_r \bigotimes\mathbf{e}_r + A_{z_1r}\mathbf{e}_{z_1} \bigotimes\mathbf{e}_r + A_{z_{1}r}\mathbf{e}_r \bigotimes\mathbf{e}_{z_1} + A_{z_1z_1}\mathbf{e}_{z_{1}} \bigotimes\mathbf{e}_{z_1} + A_{\theta\theta}\mathbf{e}_\theta \bigotimes\mathbf{e}_\theta \label{tensor}
\end{align}

where $A_{rz_{1}}=A_{z_{1}r}$ and the radial/axial $u/w$ component of the velocity field  as well as all non-zero elements of the tensor $\mathbf{A}$ are only function of $\eta$, $s$ and $\tau$ ($u=u(\eta,s,\tau)$, for example). However, care must be taken since the final expressions of $\mathbf{V}$ and $\mathbf{A}$ in $\mathbf{X}$ do depend on $\theta$ through the unit vectors given by equation (\ref{ecylindrical}) that make up the cylindrical basis.

 To obtain the equations in the bulk,  we substitute the expressions for $\mathbf{V}$ and  $\mathbf{A}$ given by (\ref{tensor}) and pressure, $p(s,\eta;\tau)$,  into  equations $(\ref{eq1})-(\ref{eq3})$, where the spatial and temporal derivatives are computed as in (\ref{chain}) taking into account that  $\mathbf{J}_\mathbf{X}$,  and $\mathbf{v}_t$  are obtained from the mapping equation (\ref{mapping3}). We arrive at a system of 14 equations in the three-dimensional space $\mathbf{X}$ and time for a set of 7 unknown  variables ($u, w, p, A_{z_1z_1}, A_{rr}, A_{z_1r}, A_{\theta\theta})$ defined in the space $[s,\eta,\theta;\tau]$. To obtain a system of equations consistent with the number of unknowns and not dependent on $\theta$ we just need to project by the cylindrical basis vectors and simplify:

\begin{align} 
\nabla \cdot \mathbf{V} &= 0, \label{eq1a}\\
  \mathbf{M}_\mathbf{X}\cdot \mathbf{e}_r&=0,\quad\mathbf{M}_\mathbf{X}\cdot \mathbf{e}_{z_1}=0 ,\label{eq2a}\\
\mathbf{e}_r\cdot\mathbf{C}_{\mathbf{X}\mathbf{X}}\cdot \mathbf{e}_r^\top&=0,\quad \mathbf{e}_{z_1}\cdot\mathbf{C}_{\mathbf{X}\mathbf{X}}\cdot\mathbf{e}_{z_1}^\top=0,\quad \mathbf{e}_\theta\cdot\mathbf{C}_{\mathbf{X}\mathbf{X}}\cdot \mathbf{e}_\theta^\top=0,\quad\mathbf{e}_{z_1}\cdot\mathbf{C}_{\mathbf{X}\mathbf{X}}\cdot \mathbf{e}_r^\top=0.\label{eq3a}
\end{align}

In addressing this problem, it is reasonable to assume that the azimuthal velocity is zero, $v=0$. This is due to the conditions of zero tangent forces in the interface and the absence of mass forces that would disrupt the axisymmetry of the problem. A test of the consistency of the method is to check that there is no need to add  the equation  $\mathbf{M}_\mathbf{X}\cdot \mathbf{e}_\theta=0$ (conservation of the azimuthal momentum) since it should be directly zero once the corresponding simplifications have been taken into account.
 The next step is to work with the equations we are imposing on the free interface. For them, and taking advantage of our three-dimensional mapping of the problem, we define two tangent vectors to the interface that are linearly independent.
 \beq
 \mathbf{X}_s=\frac{\partial\mathbf{X}_i}{\partial s}=\left[f_s\cos\theta,f_s\sin\theta,1\right],\quad \mathbf{X}_\theta=\frac{\partial\mathbf{X}_i}{\partial \theta}=\left[-f\sin\theta,f\cos\theta,0\right],
 \eeq
 
where $f_s$ is the partial derivative of f with respect to s. Therefore, our normalized tangent vectors are $\mathbf{t
}_1=\frac{\mathbf{X}_s}{|\mathbf{X}_s|}$ and  $\mathbf{t
}_2=\frac{\mathbf{X}_\theta}{|\mathbf{X}_\theta|}$  while the expression for the normal vector to the surface, $\mathbf{n}=\mathbf{t
}_2\times\mathbf{t
}_1$,  is given by
\beq
\mathbf{n} (s,\eta,\theta;\tau)=\frac{1} {\sqrt{1+fs^2}}\left[\cos\theta,\sin\theta,-f_s]\right]\label{normal}
\eeq

In the light of the above definitions and the fact that the total stress tensor in $\mathbf{X} $ has already been calculated for the bulk equations, it only remains to calculate the curvature appearing in the normal force balance on the surface. To perform this calculation, we can use the fundamental surface formulae. These formulae, together with the Christoffel symbols, have been of great importance when working with curved spaces. However, in this example we will demonstrate that with the current tools of symbolic calculus they are not necessary, given that the some operations on the surface are no more than operations carried out in three-dimensional Euclidean space and projected onto a two-dimensional space. An examination of the equation (\ref{normal}) for the normal vector reveals that, although the normal vector is only defined at the interface ($\eta = 1$), the formula makes it possible to obtain the three-dimensional spatial and temporal variation of this vector, $\mathbf{n}(x, y, z; t)$. To calculate the curvature, it is sufficient to calculate the divergence, where the spatial derivatives are calculated as in equation (\ref{chain}) and evaluated at the interface ($\eta = 1$). We finally arrive at the following expression 

\beq
 \kappa=\frac{f_s^2 - ff_{ss} + 1}{f(f_s^2 + 1)^{3/2}}.
\eeq
As anticipated, this expression is independent of $\theta$. It is widely recognised and can be derived from the fundamental formula for the curvature.
Finally, the implicit equation for the interface is
\beq
F_{surface}=r-f(z,\tau)=0, \quad r=\sqrt{x^2+y^2}\quad z=s.
\eeq
Therefore, the kinematics equation is
\beq
\frac{D F_{surface}}{Dt}=u-\frac{\partial f}{\partial \tau}-w\frac{\partial f}{\partial s}=0.  \label {kinematics}
\eeq
Equations (\ref{surfaceeq}) and (\ref{kinematics}) provides three equations for $w$, $u$ and the shape $f$ (note that $\mathbf{t}_2\cdot\mathbf{\sigma}\cdot\mathbf{n}=0$ is automatically satisfied). Neumann conditions are applied at the surface for the rest of unknowns. 
To close the problem we impose symmetric conditions at $s=z=0$ at $z=s=2\pi$ and regularity conditions at the axis $\eta=0$. 

\subsubsection{Numerical implementation}  

The two-dimensional numerical domain, defined as the interval $[0\leq s\leq 2\pi]\times [0\leq \eta\leq 1]$, is discretised in the direction of the variable $\eta$ using $n_\eta$ Chebyshev spectral collocation points.  Conversely, in the s-direction, forth-order finite differences with $n_s$ equally spaced points are employed. The results presented in the nex section were obtained using $n_s=251$, $n_\eta=10$. The temporal derivatives have been discretised using second-order backward derivatives and the problem has been  being integrated in time using an adaptive time step, $\Delta t $.

\begin{figure}
\centering
\includegraphics[width=16 cm]{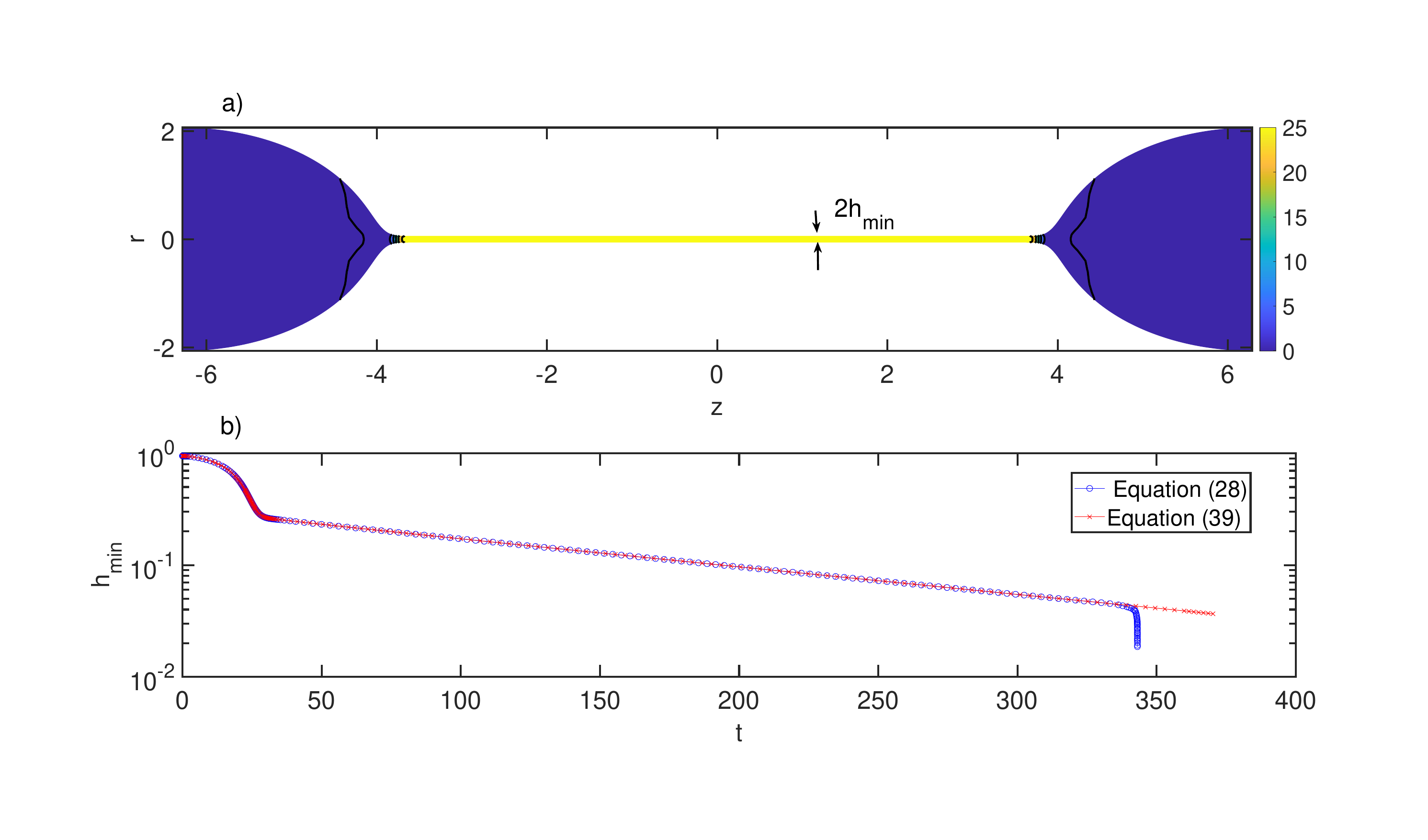}
\caption{\label{hmin}
(a) Contours of axial viscoelastic stresses, $\mathbf{\sigma}_{pz_1z_1}$ at $t=310$ and (b) $h_{min}$ as a function of time: The blue circular line was obtained using  equation (\ref{tensor}), while the red circular line was obtained using  equation (\ref{tensor1}). }
\end{figure}

\subsubsection{Results} 

We present the results for  $Oh=3.16$. $De=60$ and $S=0.25$ chosen as in reference \cite{Turkoz2018}. Figure (\ref{hmin}) (a) shows contours of  axial viscoelatic stresses $\mathbf{\sigma}_{pz_1z_1}$ at time $t=310$. We can see the formation of a uniform long thread of radius $h_{min}(t)=min(f)_s$ due to an  increase of the axial viscoelastic stresses. In (b), $h_{min}(t)$ is shown to decrease exponentially in time, with decay exponent $1/3De$,
\beq
h_{min}(t)\sim \exp\left(-\frac{t}{3De}\right).
\eeq

This example was also used to illustrate the formation of similarity solutions in the corner region between the thread and the drop\cite{Eggers_Herrada_Snoeijer_2020}. The numerical method employed was identical to that used in this study, with the exception that the authors worked directly with the viscoelastic stresses utilising the equations in axisymmetric cylindrical coordinates and introduced the logarithmic transform, as referenced in \cite{FATTAL2004281}. This allowed the authors to delay the appearance of the instability associated with such a highly hyperbolic case, which is sometimes referred to as the high Weissenberg number problem (HWNP).
Our simulations using the cylindrical basis formulation and where the conformal tensor A is integrated directly in time, show that the instability appears at an earlier stage, as illustrated the blue line in Figure (\ref{hmin}) b).  In order to explore whether the stability of the code could be improved changing the basis, the problem has been reformulated in terms of

\begin{align}
\mathbf{V} &= u^\eta\mathbf{b}_\eta + w^s\mathbf{b}_s \nonumber \\
\mathbf{A} &= A^{ss}\mathbf{b}_s \bigotimes\mathbf{b}_s + A^{s\eta}\mathbf{b}_s \bigotimes\mathbf{b}_\eta + A^{s\eta}\mathbf{b}_\eta \bigotimes\mathbf{b}_s + A^{ss}\mathbf{b}_s \bigotimes\mathbf{b}_s + A^{\theta\theta}\mathbf{b}_\theta \bigotimes\mathbf{b}_\theta\label {tensor1}
\end{align}

where
$$
\mathbf{b}_\eta=\frac{\partial\mathbf{X}}{\partial s},\quad \mathbf{b}_s=\frac{\partial\mathbf{X}}{\partial \eta},\quad \mathbf{b}_\theta=\frac{1}{r}\frac{\partial\mathbf{X}}{\partial \theta}.
$$
At this point the full potential and flexibility of the proposed method becomes evident. While the classical procedure would involve searching the literature for expressions of the operators in this basis or obtaining them by applying Christoffel symbols and then implementing them numerically, our proposal is as simple as introducing these expressions into the 3D Cartesian equations (\ref{eq1})-(\ref{eq3}) and then reducing the number of equations by scalar multiplication by an independent basis. For example, we could use the (\ref{eq1a})-(\ref{eq3a}) projection (cylindrical projection) to reduce the number of equations in the problem. The red curve in figure (\ref{hmin}) b) shows the simulation results using this basis. As can be seen, both simulations with the same discretisation produce the same results, although with the new basis the simulation takes slightly longer to become unstable.
\subsubsection{Filtering} 
\begin{figure}
\centering
\includegraphics[width=16 cm]{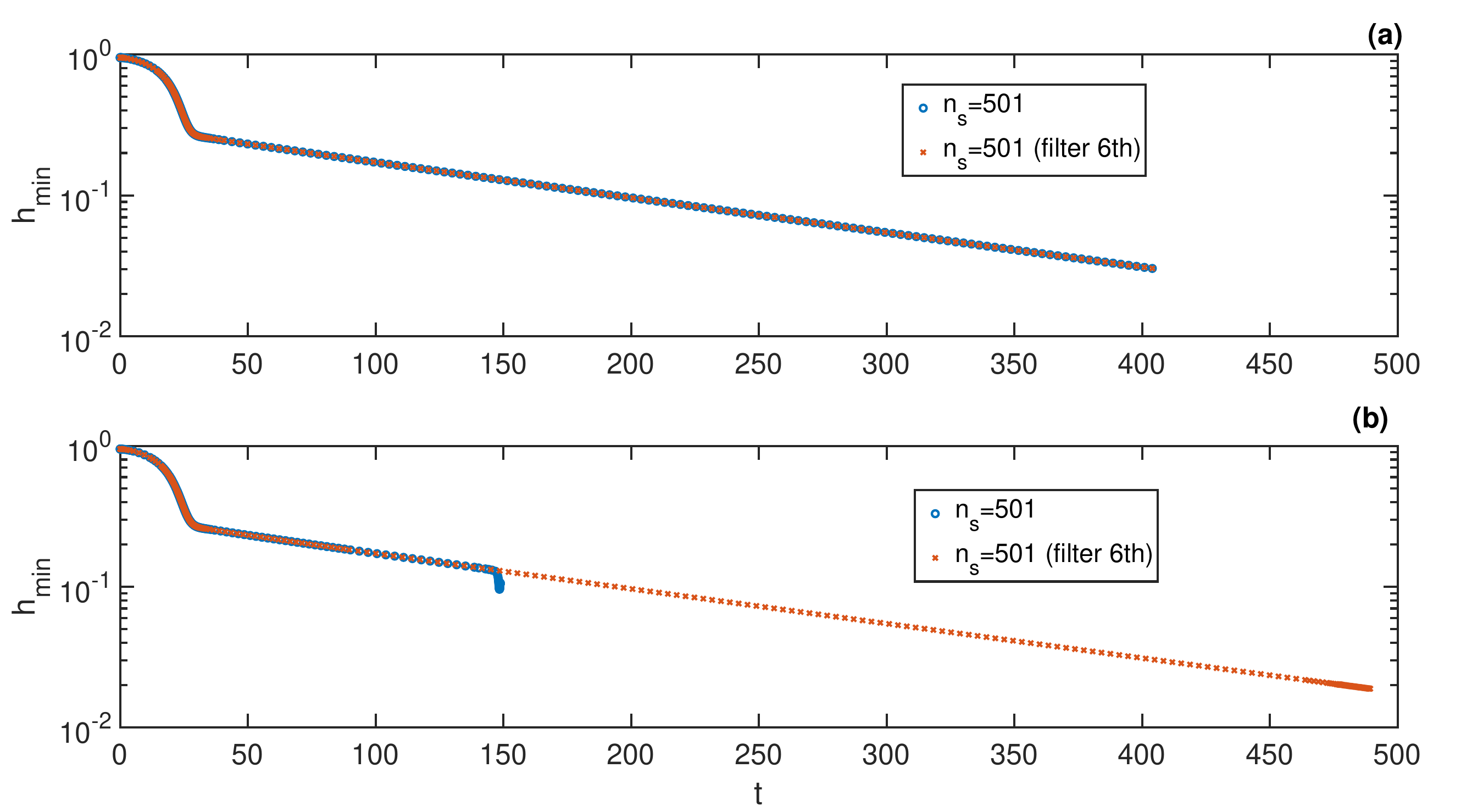}
\caption{\label{hminfilter}
$h_{min}$ as a function of time for a mesh with $n_s=501$ without filtering (circle symbols) and with filtering (cross symbols):
(a) Using basis (\ref{tensor}), (b) Using basis (\ref{tensor1}).}
\end{figure}

The numerical code make use of a fourth-order central finite differencing for discretisation in the s-direction. It is well established that the use of centred differencing in a non-staggered grid in strongly hyperbolic problems can result in code instability. This is illustrated in the figure  (\ref{hminfilter}), which depicts the results of a simulation with a finer mesh in the s-direction. It can be observed that while the use of a finer mesh with the cylindrical formulation ( figure \ref{hminfilter} (a)) results in longer times, the alternative basis formulation (figure \ref{hminfilter}(b)) leads to an earlier onset of code instability.    It is hypothesised that the discrepancy in behaviour is attributable to the fact that, while in the cylindrical formulation the shape derivatives (characterised by the $f$ function) are absent from the expressions for the velocity field and stress tensor, they are present when the alternative basis is employed. As a result, any numerical curl in the s-direction is further amplified in the alternative basis. To test this hypothesis, the simulations were repeated, with the variables in the s-direction filtered out at each time step using a sixth-order filter\cite{VISBAL2002155}. While the results remained unchanged for the cylindrical basis case, the alternative basis case exhibited a notable improvement, with the solution capable of reaching longer times than with the cylindrical basis (with the same mesh). It is hypothesised that the use of a mesh-aligned basis enables the filter to be much more effective in removing numerical waviness. The improvements obtained with this new formulation provide a basis for exploring its use in other viscoelastic fluid problems as an alternative to the use of the complicated logarithmic transformation, which is widely employed in this field of study.


\section{Conclusions}

In this paper, we introduced a novel approach for solving physical problems in curved spaces without the need for Christoffel symbols. Our method offers several key advantages:

1. **Simplification of Differential Operators:** All equations are expressed directly in the Euclidean space \(X\) and time \(t\), where differential operators are simple to compute. This eliminates the need for manually translating physical problems into specific coordinate systems, reducing errors that commonly arise from incorrectly transcribed terms or missing terms from reference texts.

2. **Flexibility in Choice of Basis:** Our method allows for flexibility in selecting the basis for vectors and tensors, as long as the chosen basis is linearly independent. This flexibility enables users to select a basis that enhances the stability of the numerical solution. While the physical results should be independent of the chosen basis, the numerical performance—such as the condition of the Jacobian matrix—can vary significantly between bases.

3. **Improved Numerical Stability:** The flexibility in basis selection allows for the optimization of the numerical method by choosing a basis that minimizes numerical errors. For example, in some cases, ill-conditioned Jacobian matrices in one basis may become better conditioned in another, leading to faster and more stable simulations.

In summary, this method simplifies both the analytical and numerical treatment of physical problems in curved spaces and offers a versatile framework that can be applied to a wide range of problems, from fluid mechanics to complex free surface flows. Future work will explore further applications of this method, particularly in the context of highly non-linear and hyperbolic systems, where numerical stability remains a challenge.

The author wishes to express his gratitude to Professors Jens Eggers, Pascual Riesco-Chueca and José María López-Herrera for their valuable comments and suggestions to improve the quality of the article. This research was supported by the Ministerio de Ciencia e Innovación (Spain), grant no. PID2022-14095OB-C21.

\appendix
\section{Derivation of the Navier-Stokes Equations in Cylindrical Coordinates using Christoffel Symbols}

In curvilinear coordinates, the position of a point in space is described by coordinates \( x^i \), where \(i = 1, 2, 3\). In cylindrical coordinates, these curvilinear coordinates are:
\[
x^1 = r, \quad x^2 = z_1, \quad x^3 = \theta,
\]

The velocity field \(\mathbf{V}\) is expressed in term of the natural basis (\ref{basis}):
\[
\mathbf{V} = u(r, \theta, z_1;t) \mathbf{b}_1 + w(r, \theta, z_1;t) \mathbf{b}_{2}+ v(r, \theta, z_1;t)/r \mathbf{b}_{3}
\]
where \(u\), \(v\), and \(w\) are the velocity components in the radial, azimuthal, and axial directions, respectively. To compute differential operators like the gradient and Laplacian in cylindrical coordinates, we must account for the metric tensor \( g_{ij} \) and its inverse \( g^{ij} \), as well as the Christoffel symbols \cite{frankel2011, nakahara2021geometry}.
The  metric tensor $\mathbf{g}$ were $g_{ij}=\mathbf{b}_i\cdot \mathbf{b}_j$, ($i=1,3$ ,$j=1,3$) is given by
 
\[
g_{ij} = \begin{pmatrix}
g_{11} & 0 & 0 \\
0 & g_{22} & 0 \\
0 & 0 & g_{33}
\end{pmatrix} = \begin{pmatrix}
1 & 0 & 0 \\
0 & 1 & 0 \\
0 & 0 & r^2
\end{pmatrix}.
\]
The inverse metric tensor \( g^{ij} \) is defined such that:
\[
g_{ik} g^{kj} = \delta_i^j,
\]
where \( \delta_i^j \) is the Kronecker delta. For cylindrical coordinates, the inverse metric tensor \( g^{ij} \) is:
\[
g^{ij} = \begin{pmatrix}
1 & 0 & 0 \\
0 &  1& 0 \\
0 & 0 & \frac{1}{r^2}
\end{pmatrix}.
\]

The Christoffel symbols of second kind \( \Gamma^k_{ij} \) are calculated from the metric tensor using the formula:
\[
\Gamma^k_{ij} = \frac{1}{2} g^{kl} \left( \frac{\partial g_{il}}{\partial x^j} + \frac{\partial g_{jl}}{\partial x^i} - \frac{\partial g_{ij}}{\partial x^l} \right).
\]
For cylindrical coordinates, the non-zero Christoffel symbols are:
\[
\Gamma^1_{33} = -r, \quad \Gamma^3_{13}  = \Gamma^3_{3 1} = \frac{1}{r}.
\]
The divergence of the velocity  field \(\mathbf{V}\) with components \( v_i \) is given by:
\[
\nabla \cdot \mathbf{V} = \left[\frac{\partial v_i}{\partial x^j}-\Gamma^l_{ji} v_l\right] g^{ij},
\]

Therefore, the continuity equation for incompressible flow in cylindrical coordinates is:
\beq
\nabla \cdot \mathbf{V}=\frac{1}{r} \frac{\partial}{\partial r}(r u) + \frac{1}{r} \frac{\partial v}{\partial \theta} + \frac{\partial w}{\partial z_1} = 0.
\eeq

In curvilinear coordinates, the gradient of pressure field p is written as:
\[
\nabla p =  \frac{\partial p}{\partial x^i}\mathbf{b}^i.
\]
where $\mathbf{b}^i$ is the contravariant basis, $\mathbf{b}^1=\mathbf{e}_r$, $\mathbf{b}^2=\mathbf{e}_{z1}$ and $\mathbf{b}^3=1/r \mathbf{e}_\theta$. Therefore
\beq
\nabla p = \frac{\partial p}{\partial r} \mathbf{e}_r + \frac{1}{r} \frac{\partial p}{\partial \theta} \mathbf{e}_\theta + \frac{\partial p}{\partial z_1} \mathbf{e}_{z_1},\label{gradientp}
\eeq

The gradient of a vector field \(v_i\) in curvilinear coordinates is written as:

 \[
   \boldsymbol{\nabla}\mathbf{V} = \left[\cfrac{\partial v_i}{\partial x^k} - \Gamma^l_{ki}~v_l\right]~\mathbf{b}^i\otimes\mathbf{b}^k
  \]

For cylindrical coordinates, using the Christoffel symbols, we obtain:

   \begin{align}
   \boldsymbol{\nabla}\mathbf{V} & = \cfrac{\partial u}{\partial r}~\mathbf{e}_r\otimes\mathbf{e}_r + 
     \cfrac{1}{r}\left(\cfrac{\partial u}{\partial \theta} - v\right)~\mathbf{e}_r\otimes\mathbf{e}_\theta + \cfrac{\partial u}{\partial z_1}~\mathbf{e}_r\otimes\mathbf{e}_{z1} \label{gradientV}\\[8pt]\nonumber
 & + \cfrac{\partial v}{\partial r}~\mathbf{e}_\theta\otimes\mathbf{e}_r + 
     \cfrac{1}{r}\left(\cfrac{\partial v}{\partial \theta} + u\right)~\mathbf{e}_\theta\otimes\mathbf{e}_\theta + \cfrac{\partial v}{\partial z_1}~\mathbf{e}_\theta\otimes\mathbf{e}_{z_1} \\[8pt]\nonumber
 & + \cfrac{\partial w}{\partial r}~\mathbf{e}_{z_1}\otimes\mathbf{e}_r + 
     \cfrac{1}{r}\cfrac{\partial w}{\partial \theta}~\mathbf{e}_{z_1}\otimes\mathbf{e}_\theta + \cfrac{\partial w}{\partial z_1}~\mathbf{e}_{z1}\otimes\mathbf{e}_{z_1}.
   \end{align} 
On the order hand, the divergence of a second order tensor $\mathbf{S}$ can be obtained from
  \[
\boldsymbol{\nabla}\cdot\boldsymbol{S}  = \left[\cfrac{\partial S_{ij}}{\partial x^k} - \Gamma^l_{ki}~S_{lj} - \Gamma^l_{kj}~S_{il}\right]~g^{ik}~\mathbf{b}^j 
  \]
 Using the above definition we can show that the divergence of a second-order tensor field, $\mathbf{S}$, in cylindrical  coordinates can be expressed as
\begin{align}
\boldsymbol{\nabla}\cdot \boldsymbol{S} & = \frac{\partial S_{rr}}{\partial r}~\mathbf{e}_r 
   + \frac{\partial S_{r\theta}}{\partial r}~\mathbf{e}_\theta
   + \frac{\partial S_{rz_1}}{\partial r}~\mathbf{e}_{z_1}  \label{divergenceS}\\[8pt] \nonumber 
 &  + \cfrac{1}{r}\left[\frac{\partial S_{r \theta}}{\partial \theta} + (S_{rr}-S_{\theta\theta})\right]~\mathbf{e}_r  +
\cfrac{1}{r}\left[\frac{\partial S_{\theta\theta}}{\partial \theta} + (S_{r\theta}+S_{\theta r})\right]~\mathbf{e}_\theta   +\cfrac{1}{r}\left[\frac{\partial S_{\theta z_1}}{\partial \theta} + S_{rz_1}\right]~\mathbf{e}_{z_1}  \nonumber \\[8pt]
 &  +
\frac{\partial S_{zr}}{\partial z_1}~\mathbf{e}_r +
\frac{\partial S_{z_1\theta}}{\partial z_1}~\mathbf{e}_\theta +
\frac{\partial S_{z_1z_1}}{\partial z_1}~\mathbf{e}_{z_1}. \nonumber 
\end{align}

To calculate the Laplacian of the velocity field we only need to apply the divergence operator given by the equation (\ref{divergenceS}) to the tensor $\boldsymbol{S}=\boldsymbol{\nabla}\mathbf{V}$ given by the equation (\ref{gradientV})
\begin{equation}
    \nabla^2\mathbf{V}=\nabla\cdot \left( \nabla\mathbf{V}\right)\label{laplaceV}
\end{equation}

Finally substituting (\ref{gradientp}), (\ref{gradientV}), and (\ref{laplaceV}) into (\ref{MF}) we arrive at the following equations

The radial momentum equation:
\[
\rho \left( \frac{\partial u}{\partial t} + u \frac{\partial u}{\partial r} + \frac{v}{r} \frac{\partial u}{\partial \theta} + w \frac{\partial u}{\partial z_1} - \frac{v^2}{r} \right) = -\frac{\partial p}{\partial r} + \mu \left( \nabla^2 u - \frac{u}{r^2} \right).
\]

The azimuthal momentum equation:
\[
\rho \left( \frac{\partial v}{\partial t} + u \frac{\partial v}{\partial r} + \frac{v}{r} \frac{\partial v}{\partial \theta} + w \frac{\partial v}{\partial z_1} + \frac{u v}{r} \right) = -\frac{1}{r} \frac{\partial p}{\partial \theta} + \mu \left( \nabla^2 v - \frac{v}{r^2} \right).
\]

The axial momentum equation:
\[
\rho \left( \frac{\partial w}{\partial t} + u \frac{\partial w}{\partial r} + \frac{v}{r} \frac{\partial w}{\partial \theta} + w \frac{\partial w}{\partial z_1} \right) = -\frac{\partial p}{\partial z_1} + \mu \nabla^2 w,
\]
where

\[
\nabla^2  = \frac{1}{r} \frac{\partial}{\partial r} \left( r \frac{\partial }{\partial r} \right) + \frac{1}{r^2} \frac{\partial^2 }{\partial \theta^2} + \frac{\partial^2 }{\partial z_1^2}.
\]

\bibliographystyle{elsarticle-num}

\end{document}